\documentclass[twocolumn,superscriptaddress,reprint]{revtex4-1}

\usepackage{amsmath}
\usepackage{amssymb}
\usepackage{graphicx}
\usepackage{color}  

\usepackage[dvipdfm]{hyperref} 
\usepackage{tabularx}

\begin{document}

\title{A polarized neutron diffraction study of the field-induced magnetization in the normal and superconducting states of Ba(Fe$_{1-x}$Co$_{x}$)$_2$As$_2$ ($x$=0.65)}

\author{C. Lester}
\affiliation{H.H. Wills Physics Laboratory, University of Bristol,
Tyndall Ave., Bristol, BS8 1TL, United Kingdom}

\author{Jiun-Haw Chu}
\affiliation{Geballe Laboratory for Advanced Materials and
Department of Applied Physics, Stanford University, Stanford, CA
94305}
\affiliation{Stanford Institute for Materials and Energy
Sciences, SLAC National Accelerator Laboratory, 2575 Sand Hill
Road, Menlo Park, CA 94025}

\author{J. G. Analytis}
\affiliation{Geballe Laboratory for Advanced Materials and
Department of Applied Physics, Stanford University, Stanford, CA
94305}
\affiliation{Stanford Institute for Materials and Energy
Sciences, SLAC National Accelerator Laboratory, 2575 Sand Hill
Road, Menlo Park, CA 94025}

\author{A. Stunault}
\affiliation{Institut Max von Laue-Paul Langevin, 38042 Grenoble,
France}

\author{I. R. Fisher}
\affiliation{Geballe Laboratory for Advanced Materials and
Department of Applied Physics, Stanford University, Stanford, CA
94305}
\affiliation{Stanford Institute for Materials and Energy
Sciences, SLAC National Accelerator Laboratory, 2575 Sand Hill
Road, Menlo Park, CA 94025}

\author{S.M. Hayden}
\email{s.hayden@bris.ac.uk}
\affiliation{H.H. Wills Physics Laboratory, University of Bristol,
Tyndall Ave., Bristol, BS8 1TL, United Kingdom}

\begin{abstract}

We use polarised neutron diffraction to study the induced magnetization density of near optimally doped Ba(Fe$_{0.935}$Co$_{0.065}$)$_2$As$_2$ ($T_C$=24~K) as a function of magnetic field ($1 \! < \! \mu_0 H \! < \! 9$~T) and temperature ($2 \! < \! T \! < \! 300$~K). The $T$-dependence of the induced moment in the superconducting state is consistent with the Yosida function, characteristic of spin-singlet pairing. The induced moment is proportional to applied field for $\mu_0 H \leq \mbox{9 T} \approx \mu_0 H_{c2}/6$. In addition to the Yosida spin-susceptibility, our results reveal a large zero-field contribution $M (H \rightarrow 0,T \rightarrow 0)/H \approx 2/3 \chi_{\mathrm{normal}}$ which does not scale with the field or number of vortices and is most likely due to the van Vleck susceptibility.   Magnetic structure factors derived from the polarization dependence of 15 Bragg reflections were used to make a maximum entropy reconstruction of the induced magnetization distribution in real space.  The magnetization is confined to the Fe atoms and the measured density distribution is in good agreement with LAPW band structure calculations which suggest that the relevant bands near the Fermi energy are of the $d_{xz/yz}$ and $d_{xy}$ type.

\end{abstract}

\pacs{74.70.Xa, 74.25.Jb, 74.25.N-, 75.25.-j}

\maketitle

\section{Introduction}

The discovery of superconductivity in iron-based materials such as LaO$_{1-x}$F$_x$FeAs \cite{Kamihara2008a} has ignited intense interest in this field. The common features of the iron-based superconductors appear to be that they are semi-metals which have electron and hole Fermi surface pockets, separated by a $(\pi,\pi)$ wavevector \cite{Guo_note}.
Experiments have demonstrated the existence of strong spin excitations with this same wavevector for superconducting compositions. It is widely believed that iron-based superconductivity is mediated by these spin excitations resulting in superconducting states such as $s_{\pm}$, where the sign of the gap changes between different Fermi surface sheets.

Of particular interest in the iron-based superconductors is the structure of the superconducting gap and the nature of the low-energy electronic states. Penetration depth \cite{Fletcher2009a} and thermal conductivity studies \cite{Reid2010a} of a number of materials (e.g. LaFePO, KFe$_2$As$_2$ and BaFe$_2$(As$_{1-x}$P$_x$)$_2$) show evidence for low-energy quasiparticle excitations which could be due to nodes in the superconducting order parameter.

Ba(Fe$_{1-x}$Co$_x$)$_2$As$_2$ is a good system to study since it is possible to grow large single crystals
with homogeneous doping. It has been studied widely by various probes including angle-resolved photoemission spectroscopy (ARPES) \cite{Terashima2009a}, scanning tunnel microscopy (STM) \cite{Yin2009a}, penetration depth \cite{Gordon2010a,Luan2011a}, $\mu$SR \cite{Williams2009a}, heat capacity~\cite{Gang2010a,Gofryk2011a} and thermal conductivity \cite{Reid2010a}. Even in this single system, different gap characters have been proposed as a function of doping, including fully and nodally gapped structures \cite{Terashima2009a, Yin2009a, Gordon2010a, Williams2009a, Gang2010a,Gofryk2011a,Reid2010a}.

Here we use half-polarized neutron diffraction to measure the susceptibility and induced magnetization in the normal and superconducting states of near optically doped Ba(Fe$_{1-x}$Co$_x$)$_2$As$_2$ ($x=0.065$). Our measurements shed light on the electronic structure and the nature of the low energy electronic states in both phases.  By measuring the flipping ratios of a number of Bragg peaks, we are able to extract the spatial Fourier components of the induced magnetization density $\mathbf{M(r)}$. In a metal, this provides information about the electronic states near the Fermi energy. We compare our results with a band structure calculation. In addition to measuring the $\mathbf{M(r)}$, we also made a detailed study of the temperature and magnetic field dependence of the induced magnetization by measuring a single Bragg peak in detail.

The paper is organised as follows. In Sec.~\ref{Sec:Background} we introduce the polarized beam method used in our experiment and report our unpolarized structural refinement of Ba(Fe$_{1-x}$Co$_x$)$_2$As$_2$ ($x=0.065$). In Sec.~\ref{Sec:MagDens} we report our determination of the real space magnetization density $\mathbf{M(r)}$ induced by an applied magnetic field. We also present the results of a LAPW calculation of the magnetization density distribution. In Sec.~\ref{Sec:TBdep}, we report measurements of the induced magnetization \textit{in the superconducting state} as a function of magnetic field and temperature. We discuss the significance of our observations with respect to the superconductivity in Ba(Fe$_{1-x}$Co$_x$)$_2$As$_2$ ($x=0.065$) and other experimental results. This is followed by a summary of our conclusions.

\section{Background}
\label{Sec:Background}
\subsection{Polarized Neutron Diffraction Studies of the Induced Magnetization }

Polarized neutron scattering experiments can directly measure the real-space magnetization density $\mathbf{M}(\mathbf{r})$ in the unit cell, induced by a large magnetic field $\mu_0 H$. Due to the periodic crystal structure, the applied magnetic field induces a magnetization density with spatial Fourier components $\mathbf{M} (\mathbf{G})$, where $\mathbf{G}$ are the reciprocal lattice vectors, such that

\begin{equation}
\label{Eq:M_r} \bf{M(r)} = \frac{1}{\nu_{0}}
\sum_{\bf{G}}{M(G)}\rm{exp(-}\it{i}\bf{G}\cdot\bf{r})
\end{equation}
and $\nu_{0}$ is the volume of the unit cell. The Fourier components of the magnetization density are given by
\begin{equation}
\label{Eq:M_G} \mathbf{M(G)} = \int_{\mbox{unit cell}} \mathbf{M(r)} \exp(i \mathbf{G
\cdot r})\rm{d}\mathbf{r}.
\end{equation}

Neutrons interact with matter through the strong nuclear force and electromagnetic interaction. For neutrons with initial and final spin polarisations $\mathbf{\sigma}_i$ and $\mathbf{\sigma}_f$, the total scattering cross section is

\begin{eqnarray}
\label{Eq:xsection}
\left( \frac{d \sigma}{d \Omega} \right)_{\sigma_{i} \rightarrow \sigma_{f}}
 & \propto &
\left| \langle \sigma_{i} |
\frac{\gamma r_{0}}{2 \mu_{B}}
\bbox{\sigma} \cdot \mathbf{\hat{G}} \times
\left\{ \mathbf{M(G)} \times
\mathbf{\hat{G}} \right\} \right.  \nonumber \\
& + &
\left. F_{N}(\mathbf{G})
| \sigma_{f} \rangle \right|^{2},
\end{eqnarray}
where $\gamma r_0=5.36\times 10^{-15}$~m and $F_N(\mathbf{G})$ is the nuclear structure factor. The sign of the first (magnetic) term in  Eq.~\ref{Eq:xsection} can be changed by reversing the incident neutron polarization. Thus we are able to isolate the interference term between the nuclear and magnetic scattering. In this experiment we measure the flipping ratio $R$, defined as the ratio of the cross-sections with neutrons parallel and anti-parallel to the applied magnetic field.   Because the induced moment
is small, the experiment is carried out in the limit
$(\gamma r_{0}/2 \mu_{B}) M(\mathbf{G})/F_N(\mathbf{G})\ll 1$. In this limit, the flipping ratio derived from Eq.~\ref{Eq:xsection} is,
\begin{equation}
\label{Eq:Flipping}
R  =  \frac{|F_{N}(\mathbf{G})-
(\gamma r_{0} /2 \mu_{B}) {M}(\mathbf{G})|^{2}}
{|F_{N}(\mathbf{G})+
(\gamma r_{0}/2 \mu_{B}) {M}(\mathbf{G})|^{2}}
 \approx
1 - \frac{2 \gamma r_{0}}{\mu_{B}}
\frac{{M}(\mathbf{G})}
{F_{N}(\mathbf{G})}.
\end{equation}
As the nuclear structure factors $F_{N}(\mathbf{G})$ are known from the
crystal structure, Eq.~\ref{Eq:Flipping} directly gives $M({\bf G})$.

\subsection{Experimental Details}

The Ba(Fe$_{1-x}$Co$_x$)$_2$As$_2$ ($x=0.065$) single crystal used in this study was prepared by a self-flux
method~\cite{Chu2009a}, had approximate dimensions 6~$\times$~1.5~$\times$~0.2~mm$^3$ and a mass of $\sim$1.8~mg.
Similar samples were used in our previous studies \cite{Lester2009a,Lester2010a}. Resistivity and magnetization measurements on crystals from the same batch identified the superconducting transition temperature $T_c$(onset)=24~K and showed no evidence of magnetic order down to 2 K.
The bulk susceptibility in a 5~T field measured using a SQUID magnetometer was $\chi_{ab}=1.22 \times 10^{-3}$~$\mu_B$~T$^{-1}$~f.u.$^{-1}$. We note that the expected upper critical field applied in the $ab$ plane for this composition is $H_{c2,ab}=$55~T \cite{Kano2009a}.

Neutron scattering experiments were performed at the Institut Laue-Langevin, Grenoble, France. An initial unpolarized structural refinement was performed using the 4-circle D9 spectrometer with $\lambda=0.837(1)$~\AA. The results are shown in Table~\ref{table:D9}.
\begin{table}
\begin{center}
\begin{tabular}{lcccc} 
\hline\hline                        
Atom & \multicolumn{2}{c}{Position in I4/mmm} & $z$& B~(\AA$^2$) \\

\hline                  
Ba     & 2a & $(0\ 0\ 0)$    &           &  0.023(16)                  \\ 
Fe/Co  & 4d & $(\frac{1}{2}\ 0\ \frac{1}{4})$ &           &  0.110(10) \\
As     & 4e & $(0\ 0\ z)$           &       $0.35352$ &  0.083(11)     \\
\hline
\multicolumn{5}{c}{$a=3.952(2)\; \mbox{\AA}, c = 12.911(15)\; \mbox{\AA}$} \\
\multicolumn{5}{c}{$g$=22.4(1.2) rad$^{-1}$} \\
\hline \hline 
\end{tabular}
\caption{Structural parameters of Ba(Fe$_{1-x}$Co$_x$)$_2$As$_2$ ($x=0.065$). Parameters are obtained from least-squares refinement of integrated intensities measured at $T=30$~K on D9. The $B$-factor is related to the mean squared atomic displacement $\langle u^2 \rangle $ by $B=8 \pi^2 \langle u^2 \rangle$. $g$ is the width parameter of the mosaic distribution \cite{Becker1974a,mosaic_note}. \label{table:D9}}

\end{center}
\end{table}
Polarized beam measurements of the flipping ratio were made on the D3 spectrometer. The sample was mounted on a thin aluminium post with the $[1\bar{1}0]$ direction vertical and parallel to the applied field. Data was collected with an incident wavelength $\lambda=0.825$~\AA\ and a 0.5~mm Er filter (to reduce higher order contamination in the incident beam). Flipping ratios for equivalent reflections were collected, averaged and corrected for the finite beam polarization and extinction effects. The sample was cooled through $T_c$ at each field measured when collecting data in the superconducting state.

\section{Induced Magnetization Distribution}
\label{Sec:MagDens}
\subsection{Results}

\begin{figure}
\begin{center}
\includegraphics[width=0.95\linewidth]{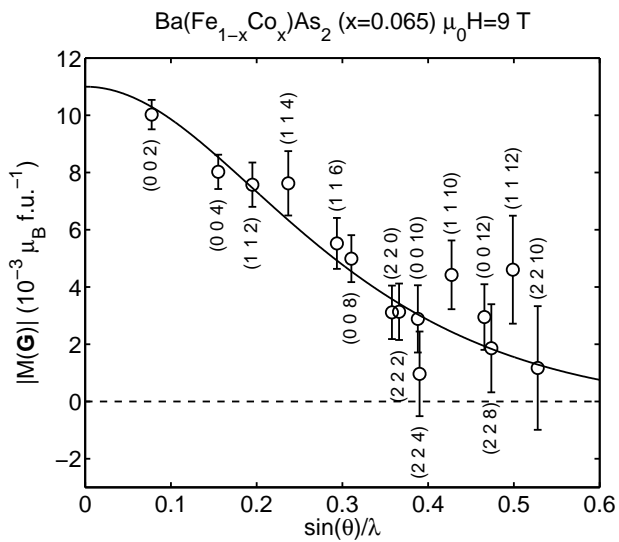}
\end{center}
\caption{The magnetic structure factors $|M(\mathbf{G})|$
measured for a 9~T field applied along the $[1\bar{1}0]$ direction. The solid line is the
Fe$^{2+}$ form factor \cite{Brown1992} scaled to the measured susceptibility. } \label{Fig:FormFactor}
\end{figure}

\begin{center}
\begin{table}
\begin{tabular}{rrrcrrr}
\hline\hline                        %
$h$ & $k$ & $l$ & \multicolumn{1}{c}{$\ \ \sin\theta/\lambda\ \ $} &   \multicolumn{1}{c}{$(1-R)$$\times$$10^3$} &
\multicolumn{1}{c}{$F_N(\mathbf{G})$} & \multicolumn{1}{c}{$M(\mathbf{G})$} \\
  &  &   & \multicolumn{1}{c}{(\AA$^{-1}$)} &   &
\multicolumn{1}{c}{(fm f.u.$^{-1}$)} & \multicolumn{1}{c}{(m$\mu_B$ f.u.$^{-1}$)} \\
\hline                  
2 & 2 & 0  & 0.3611  &  0.83 $\pm$ 0.3  &  36.16  &   3.1 $\pm$ 0.9    \\
0 & 0 & 2  & 0.0780  &  4.98 $\pm$ 0.3  & -16.88  & -10.0 $\pm$ 0.5    \\
1 & 1 & 2  & 0.1967  &  3.75 $\pm$ 0.5  &  19.93  &   7.5 $\pm$ 0.8    \\
2 & 2 & 2  & 0.3694  &  1.90 $\pm$ 0.6  & -16.54  &  -3.1 $\pm$ 1.0    \\
0 & 0 & 4  & 0.1560  &  6.49 $\pm$ 0.5  &  12.19  &   8.0 $\pm$ 0.6    \\
1 & 1 & 4  & 0.2386  &  3.08 $\pm$ 0.5  & -24.46  &  -7.6 $\pm$ 1.1    \\
2 & 2 & 4  & 0.3933  &  0.8  $\pm$ 1.3  &  12.04  &   0.9 $\pm$ 1.5    \\
1 & 1 & 6  & 0.2956  &  1.61 $\pm$ 0.2  &  32.68  &   5.5 $\pm$ 0.9    \\
0 & 0 & 8  & 0.3120  &  1.58 $\pm$ 0.4  &  29.32  &   4.9 $\pm$ 0.8    \\
2 & 2 & 8  & 0.4772  &  1.18 $\pm$ 0.7  &  28.90  &   1.8 $\pm$ 1.5    \\
0 & 0 & 10 & 0.3900  &  1.12 $\pm$ 0.5  & -25.57  &  -2.8 $\pm$ 1.2    \\
1 & 1 & 10 & 0.4298  &  4.52 $\pm$ 1.2  &  10.42  &   4.4 $\pm$ 1.2    \\
2 & 2 & 10 & 0.5315  &  3.60 $\pm$ 2.9  & -25.11  &  -1.1 $\pm$ 2.2    \\
0 & 0 & 12 & 0.4681  &  0.72 $\pm$ 0.6  &  23.42  &   2.9 $\pm$ 1.2    \\
1 & 1 & 12 & 0.5017  &  3.29 $\pm$ 4.5  & -12.16  &  -4.6 $\pm$ 1.9    \\
\hline\hline
\end{tabular}

\caption{For each Bragg reflection $(hkl)$, the table shows: $\sin\theta/\lambda$, the measured flipping ratio $R$, structure factor $F_N(\mathbf{G})$ calculated from the structure in Table~\ref{table:D9}, and the determined magnetic structure factor $M(\mathbf{G})$ \cite{SF_sign}. Data collected at $T=30$~K and $\mu_0 H=9$~T.
\label{table:D3}}
\end{table}
\end{center}
We measured the magnetization induced by a 9~T magnetic field applied along the $[1\bar{1}0]$ direction in the normal state at $T$=30~K. Table \ref{table:D3} shows the measured flipping ratios under these conditions and the extracted Fourier components of the magnetization density.  Fig.~\ref{Fig:FormFactor} shows $|M(\mathbf{G})|$ plotted against $\sin \theta /\lambda = |\mathbf{G}|/4 \pi$. If Fe were the only magnetic atom, then the sign of $M(\mathbf{G})$ is determined by the geometric structure factor for the Fe atoms and $|M(\mathbf{G})|$ is the effective form factor of the Fe atom. The solid line in Fig.~\ref{Fig:FormFactor} is the standard calculated isotropic atomic form factor for Fe$^{2+}$ \cite{Brown1992}.  Deviations from an isotropic form factor are expected at larger $\theta$ or $|\mathbf{G}|$.
Our results are in broad agreement with a recent study of Ba(Fe$_{1-x}$Co$_x$)$_2$As$_2$ $(x=0.066)$ \cite{Prokes2011a}. The main differences between the present data and the data presented in Ref.~\onlinecite{Prokes2011a} are: (i) the present study has higher statistical accuracy (ii) the value of $M(\mathbf{G})$ for $\mathbf{G}$=(002) is larger relative to the other $\mathbf{G}$ points in the present study. Our extinction model shows that the (002) peak has the largest extinction correction factor of 1.26.

\begin{figure}
\begin{center}
\includegraphics[width=0.95\linewidth]{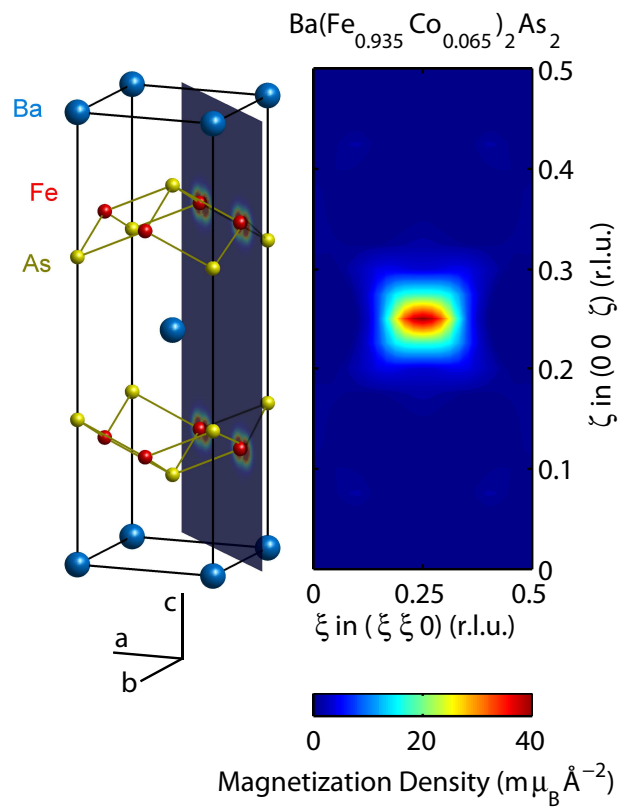}
\end{center}
\caption{(left) The conventional tetragonal unit cell of Ba(Fe,Co)$_2$As$_2$. We measure the magnetization density integrated perpendicular to the $(110)$-type plane shown. The magnetization density shown is the result of the VCA calculation shown in Fig.~\ref{Fig:MagDensity}. (right) Projected magnetization distribution reconstructed from the experimental data in Table.~\ref{table:D3}. The reconstructed magnetization map shows $\frac{1}{4}$ of the area of the plane shown in the left panel.}
\label{Fig:UnitCell}
\end{figure}

We used the maximum entropy method (MaxEnt) \cite{Skilling1989a,Papoular1990a,Gull1989a,Brown2010a} to make a model-free reconstruction of the magnetization density in real space. Flipping ratios for reflections of the type $(hhl)$ were collected, this allowed the reconstruction of the magnetization density projected down the $[1\bar{1}0]$ direction onto the $(110)$ plane as illustrated in the left panel of Fig.~\ref{Fig:UnitCell}.  The result of the reconstruction is shown in the right panel of Fig.~\ref{Fig:UnitCell}.  As expected, the magnetization density is localized mostly on the Fe atoms. The magnetization ``cloud'' appears to be slightly extended along the $\langle 110 \rangle$ directions. Our results are in broad agreement with Ref.~\cite{Prokes2011a}, however we observe no significant magnetization density on the Ba sites.  A recent study of the paramagnetic state of the parent antiferromagnet BaFe$_2$As$_2$ \cite{Brown2010a} is also broadly consistent with our results. The main difference in the BaFe$_2$As$_2$ case is that the magnetization extends more towards the As atoms.

\subsection{Electronic Structure Calculations}

Induced form factor measurements have been widely used to determine the nature of the electrons responsible for paramagnetism in solids. In metals, the induced magnetization arises from a redistribution of electrons between up and down states near the Fermi energy. Thus, we probe the nature of the electronic wavefunctions for states near the Fermi energy. In order to understand our results further, we have carried out electronic structure calculations using the WEIN2k package \cite{Blaha2001a}.

The linear augmented plane-wave (LAPW) method \cite{Singh2006a} was used to obtain the electronic structure and spin density of Ba(Fe$_{1-x}$Co$_x$)$_2$As$_2$ ($x=0.065$). We used a full-potential LAPW method with the generalized gradient approximation (GGA). In the case of doped compositions, we used the virtual crystal approximation (VCA) \cite{Nordheim1931a}. We used the lattice parameters and atom position for the Ba(Fe$_{1-x}$Co$_x$)$_2$As$_2$ $(x=0.065)$ structure shown in Table.~\ref{table:D3} for both compositions. The muffin-tin radii were chosen to be 2.5, 2.37 and 2.11 atomic units (a.u.) for Ba, Fe and As respectively, with the quantity $R_{\mathrm{MT}}K_{\mathrm{max}}$ set to 7, where $R_{\mathrm{MT}}$ is the smallest muffin-tin radius and $K_{\mathrm{max}}$ is the plane wave cutoff. For integrations we used 726 $k$-points in the irreducible Brillouin zone. In the LAPW method, the charge (spin) density is represented by a plane wave expansion in the interstitial region and as a combination of radial functions times spherical harmonics inside the muffin-tin spheres.
\begin{figure*}
\begin{center}
\includegraphics[width=0.95\linewidth]{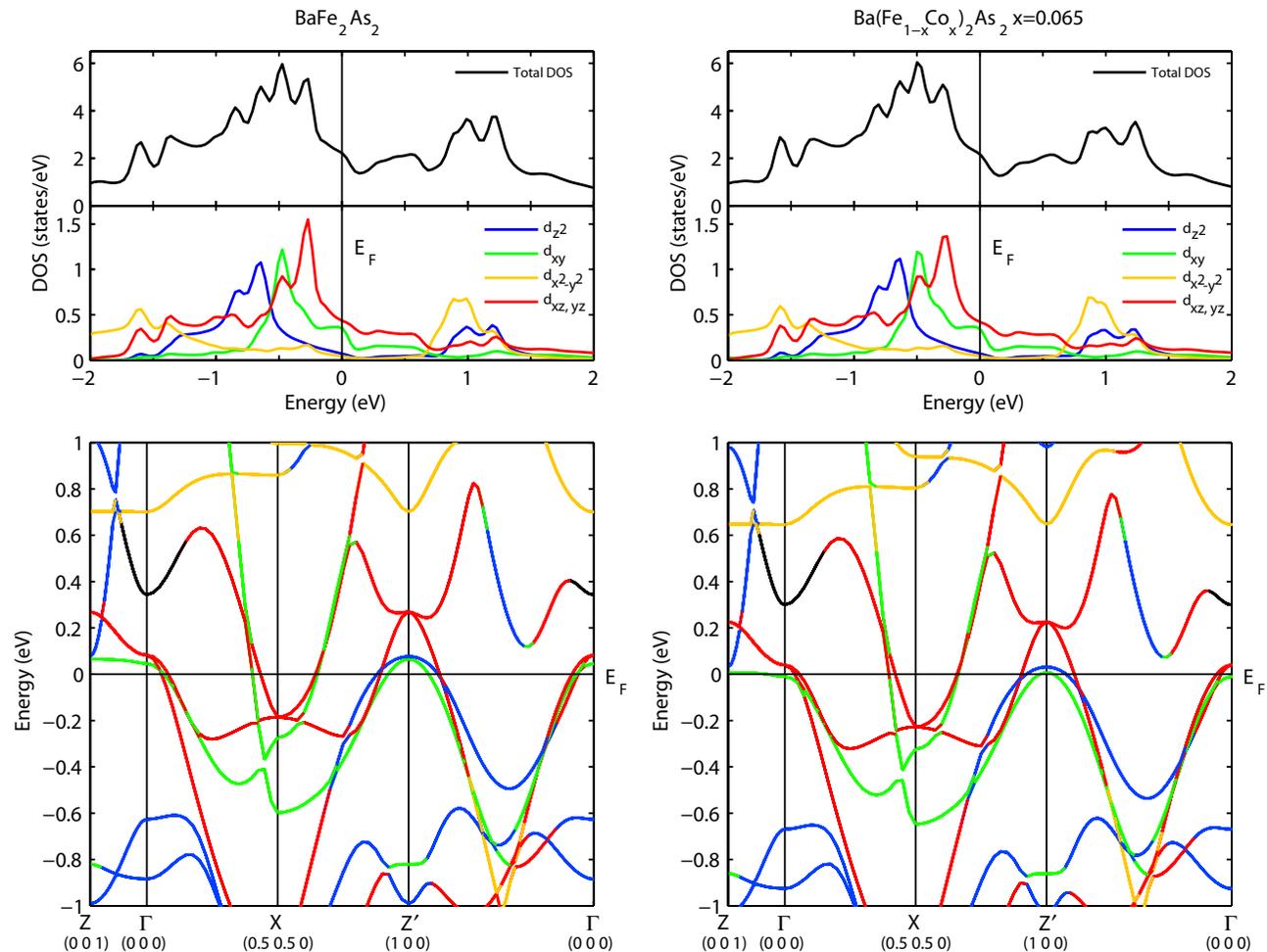}
\end{center}
\caption{(Color online) (bottom panels) The linear augmented plane-wave (LAPW) bandstructure calculations for Ba(Fe$_{1-x}$Co$_x$)$_2$As$_2$  for $x$=0 (left) and $x$=0.065 (right) in the virtual crystal approximation (VCA). Colors indicate the dominant orbital of the $\textbf{k}$ state. (top panels) Calculated electronic density of states.}
\label{Fig:BandStructure}
\end{figure*}
Fig.~\ref{Fig:BandStructure} shows the bandstructure calculated in the tetragonal phase for Ba(Fe$_{1-x}$Co$_x$)$_2$As$_2$ for $x$=0 and $x$=0.065.  Our calculations generally agree with others in the literature \cite{Singh2008a,Thirupathaiah2010a,Andersen2011a,Colonna2011a}, in particular, they show that the states near the Fermi energy are predominately of $\displaystyle{d_{xz,yz}}$ and ${\displaystyle d_{xy}}$ character.

\begin{figure}
\begin{center}
\includegraphics[width=0.95\linewidth]{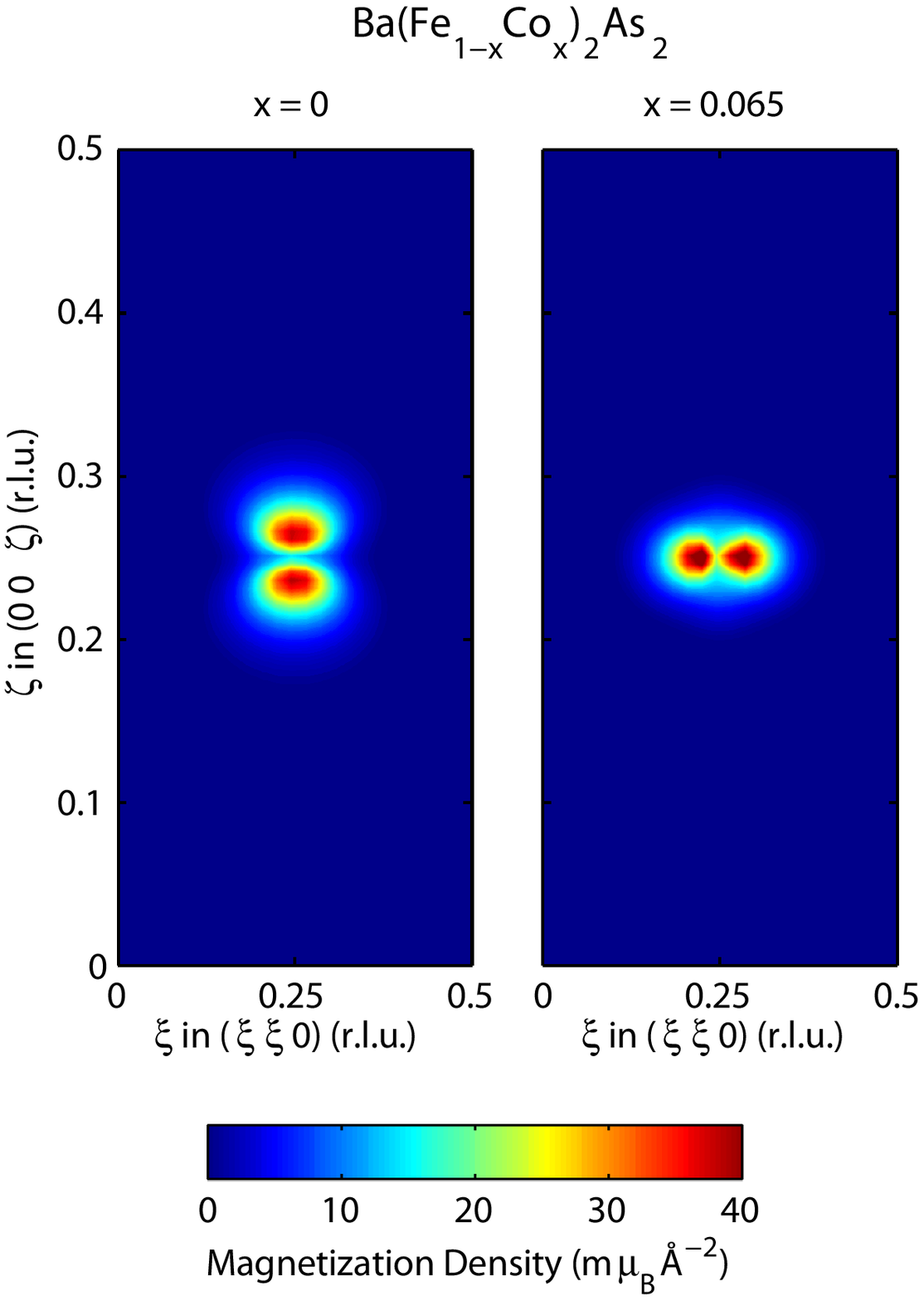}
\end{center}
\caption{Magnetization density of Ba(Fe$_{1-x}$Co$_x$)$_2$As$_2$ calculated using the LAPW method and VCA approximation for $x=0$ (left panel) and $x=0.065$ (right panel). Calculations were carried out with a small fixed ferromagnetic moment to mimic the effect of an applied field. \label{Fig:MagDensity}}
\end{figure}
Spin polarized calculations were carried out in the tetragonal state with the ferromagnetic magnetization constrained to be 0.01~$\mu_{B}$ per unit cell to mimic the effect of an applied magnetic field. The results were then scaled to the measured magnetization at $\mu_0 H=$9~T for comparison purposes. The results are shown in Fig.~\ref{Fig:MagDensity}. The $x=0.065$ calculation shows reasonable agreement with our maximum entropy reconstruction in that the magnetization density is extended along the $\langle 110 \rangle$ direction.  However, the reconstruction from our experimental data does not show the two maxima along $(\xi,\xi,1/2)$ predicted in the calculation. This is presumably because our data have insufficient Fourier components to resolve these features. We also carried out a calculation for BaFe$_2$As$_2$ ($x=0$) in the paramagnetic tetragonal state with the same structural parameters (but no Co potential) to demonstrate the sensitivity of the magnetization distribution to the bandstructure. Notice that the change in electronic structure between $x=0$ and $x=0.065$ causes a rotation of the calculated pattern in Fig.~\ref{Fig:MagDensity}.

\section{Temperature and Field Dependence of the Magnetization}
\label{Sec:TBdep}
\begin{figure}
\begin{center}
\includegraphics[width=0.85\linewidth]{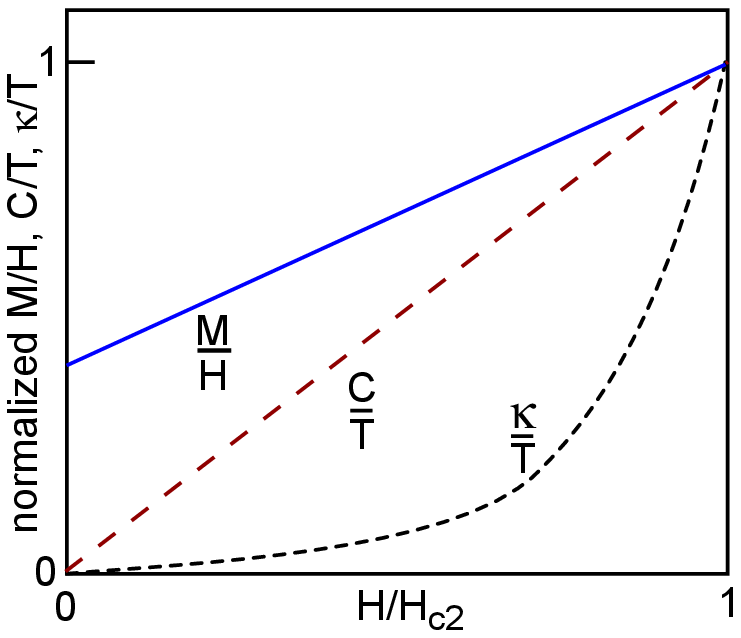}
\end{center}
\caption{Schematic field dependence of the low temperature magnetization $M$, specific heat $C$, and thermal conductivity $\kappa$ for an $s$-wave superconductor. The behavior for $C$ and $\kappa$ are based on V$_3$Si \cite{Boaknin2003a}.} \label{Fig:MCK_comparison}
\end{figure}

The measurement of the field and temperature dependence of bulk magnetization $M(H,T)$ in the mixed state of a superconductor provides information about the nature of the superconductive pairing. Thermal conductivity $\kappa(H,T,\theta)$ \cite{Reid2010a} and specific heat measurements $C=\gamma(H) T$ \cite{Moler1997a} in the mixed state have been extremely useful in probing the low energy quasi particles and identified the gap structure of a number of superconductors. In particular, the field dependence of the electronic contributions to $\kappa$ and $C$ in the $T\rightarrow0$ limit is sensitive to the symmetry of the superconducting gap function \cite{Shakeripour2009a} (see Fig. \ref{Fig:MCK_comparison}).
Complementary information is contained in $M(H,T)$.  However, studies of the bulk magnetization in the mixed state of superconductors are not possible by conventional means, e.g. SQUID magnetometery, because of the presence of a large diamagnetic contribution.  Polarized neutron diffraction and NMR Knight shift measurements \cite{Ning2008a,Oh2011a} can be used to make magnetization measurements in the mixed state. The polarized neutron diffraction technique used here is unique because it directly measures the total magnetization including spin and orbital contributions.  This technique was first used by Shull and Wedgwood in 1963 to study V$_3$Si~\cite{Shull1966a} and has subsequently been applied to such varied superconductors as UPt$_3$ \cite{Stassis1986a}, YBa$_2$Cu$_3$O$_{7-\delta}$ \cite{Boucherle1993a} and Sr$_2$RuO$_4$ \cite{Duffy2000a}.

\subsection{Results}
\begin{figure}
\begin{center}
\includegraphics[width=0.85\linewidth]{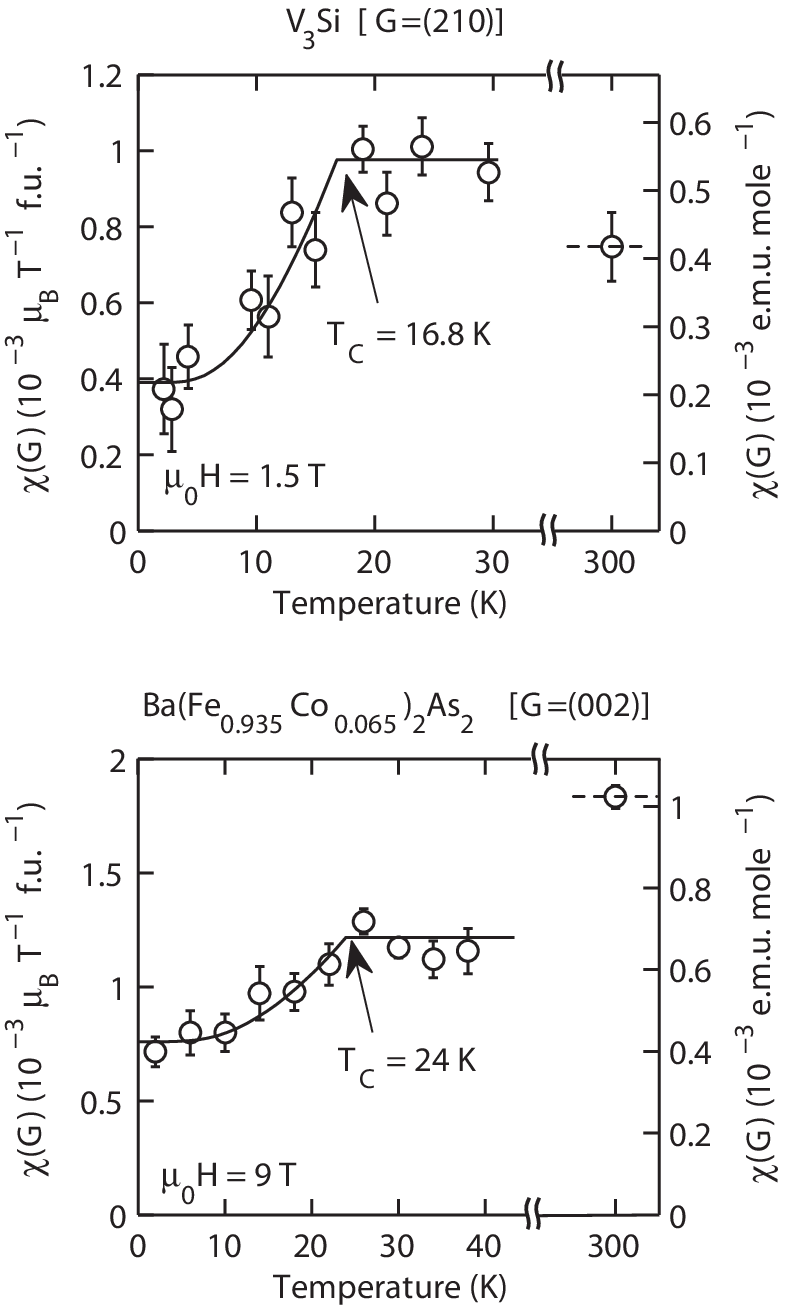}
\end{center}
\caption{The temperature dependence of the susceptibility and
induced moment of V$_3$Si~\cite{Shull1966a} (top panel) and
Ba(Fe$_{1-x}$Co$_x$)$_2$As$_2$ $x$=0.065 (bottom panel) measured using polarised
neutron scattering. The solid lines are the Yosida behaviour expected for
a singlet order parameter.} \label{Fig:TDep}
\end{figure}
\begin{figure}
\begin{center}
\includegraphics[width=0.85\linewidth]{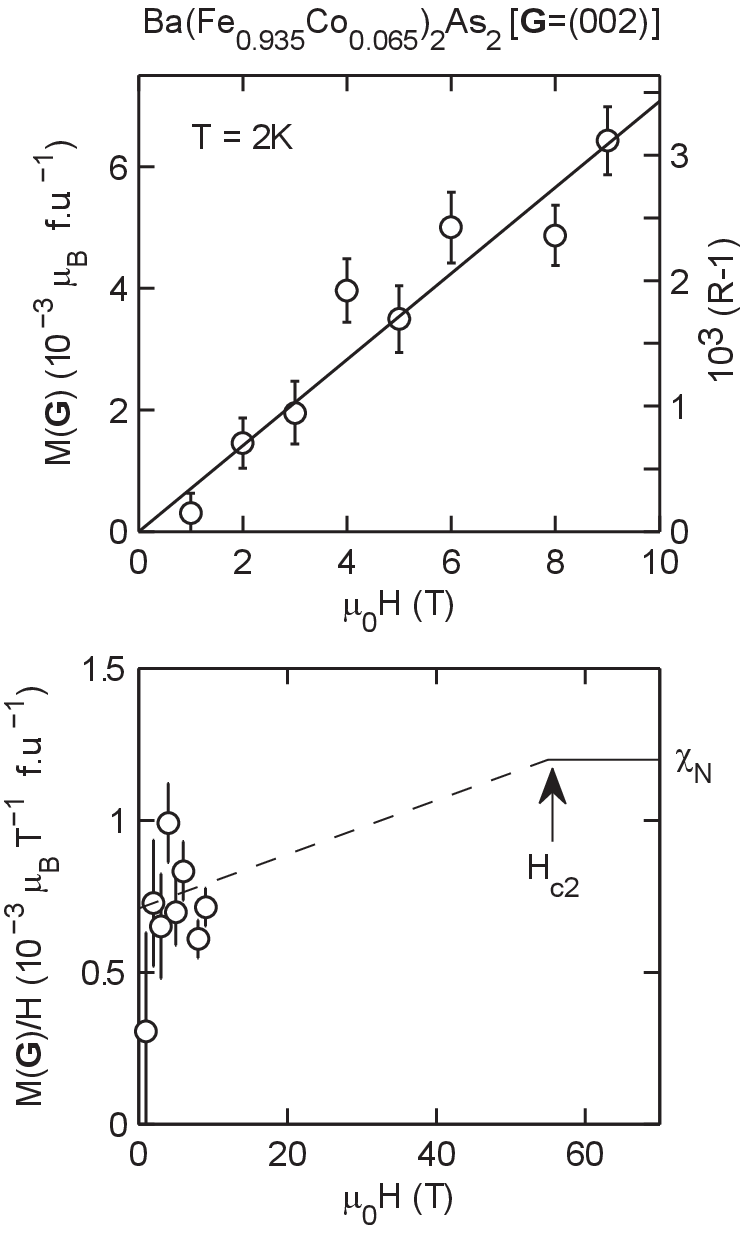}
\end{center}
\caption{(Top panel) Field dependence of the induced magnetization $M(\mathbf{G})$ of Ba(Fe$_{1-x}$Co$_x$)$_2$As$_2$ $x$=0.065 measured at $T=2$~K. $R$ is the corresponding flipping ratio. (Bottom panel) $M(\mathbf{G})/H$. The dashed line corresponds to the approximation $M_{\mathrm{spin}} \propto H^2$. \label{Fig:HDep}}
\end{figure}
We used the (002) Bragg peak to make our measurement since this requires little correction to give the $\mathbf{G}=0$ magnetization, $M(\mathbf{G}=(002))/M(\mathbf{G}=0)=0.979$ for an isotropic Fe$^{2+}$ ion.  Fig.~\ref{Fig:TDep} shows the temperature dependence the induced magnetization of Ba(Fe$_{1-x}$Co$_x$)$_2$As$_2$ ($x$=0.065) for $\mu_0 H=9$~T.   The temperature dependence shows the characteristic Yosida dependence expected for singlet pairing \cite{Yosida1958a}, this is consistent with measurements of the NMR Knight shift in the same compound \cite{Ning2008a,Oh2011a}. There is a large susceptibility in the $T\rightarrow0$ limit. A large residual contribution is also observed in V$_3$Si \cite{Shull1966a} and this has been attributed to the van Vleck contribution \cite{Clogston1962a}. The presence of large van Vleck contribution has also been inferred \cite{Grafe2008a,Graser2009a} from NMR knight shift measurements on other iron-based superconductors.
Fig.~\ref{Fig:HDep}(top panel) shows the field dependence of the induced magnetization. $M(\mathbf{G}) \propto H$ over the field range of the present experiment ($\mu_0 H < 9$~T). When we plot the susceptibility $M(\mathbf{G})/H$ [see Fig.~\ref{Fig:HDep}(bottom panel)], we find that the value as $T \rightarrow 0$ and $H \rightarrow 0$ is about $\frac{2}{3}$ of the normal state value measured at $T_c$ in the present experiment.

\subsection{Interpretation}

We measured the static response at finite wavevector $\mathbf{G}$ to an (approximately) uniform magnetic field, that is, $\chi(\mathbf{G}) \equiv \chi(\mathbf{G},0)=M({\mathbf{G}})/H(\mathbf{G}=0)$ \cite{M_sign,SF_sign}. The atomic nature of solids means that the induced magnetization $M(\mathbf{r})$ is spatially modulated on an atomic scale (see Fig.~\ref{Fig:UnitCell}). Neutrons diffract from this modulation.  In the mixed state of a superconductor, there is an additional diamagnetic magnetization (which gives rise to the vortex lattice), which is not detected in the present experiment. The signal from the vortex lattice is only present at small wavevectors (scattering angles) and can be studied by neutrons using small angle scattering techniques.    In the following discussion we do not include the superconducting diamagnetic response.

The magnetic susceptibility in $d$-band metals has several components: atomic diamagnetic, van Vleck (``orbital'' or ``interband'') and spin. The atomic diamagnetic contribution in BaFe$_2$As$_2$ has been estimated \cite{Brown2010a} to be small and is neglected here.   Only the spin contribution is expected to be suppressed by singlet Cooper pairing, thus we write the spatially averaged induced magnetization in the mixed state as:
\begin{equation}
M= \chi_{\mathrm{orb}} H + M_{\mathrm{spin}}(T,H),
\end{equation}
where $\chi_{\mathrm{orb}}$ is the orbital susceptibility and $M_{\mathrm{spin}}$ is the spin magnetization.  In an $s$-wave superconductor the density of states due to the introduction of vortices is $\propto N_F H/H_{c2}$ \cite{Volovik1993a,Caroli1964a}.  Thus the spin magnetization should vary as $M_{\mathrm{spin}}(T \rightarrow 0,H) \propto H^2$. The temperature dependence is given by the Yosida function \cite{Yosida1958a},
$M_{\mathrm{spin}}(T,H \rightarrow 0) \propto Y(T)$.  The present experiments (Fig.~\ref{Fig:HDep}) show that the differential susceptibility $dM/dH$ has a large finite value in the $H\rightarrow0$ limit and  $dM/dH \approx$ constant for $\mu_0 H < 9$~T. This is consistent with the finite $\chi(T\rightarrow 0,H \rightarrow 0)$ response being due to a van Vleck contribution.  It should be noted that there is also NMR evidence for a residual susceptibility in BaFe$_{2}$(As$_{0.67}$P$_{0.33}$)$_2$ \cite{Nakai2010a} and BaFe(Fe$_{0.93}$Co$_{0.07}$)$_{2}$As$_{2}$ \cite{Oh2011a}.  Specific heat measurements also suggest that there can be a sizeable residual quasiparticle density of states in BaFe(Fe$_{1-x}$Co$_{x}$)$_{2}$As$_{2}$
\cite{Gang2010a,Gofryk2011a,Residual_gam}. We cannot rule out the possibility that this is related to the residual susceptibility observed by neutron scattering.

There has been considerable debate about the nature of the superconducting gap in Ba(Fe$_{1-x}$Co$_{x}$)$_{2}$As$_{2}$. In principle, a detailed measurement of the $T$-dependence of the induced moment could be used to distinguish between different models for the gap. Unfortunately, the statistical noise in the present data is relatively high.  Thus we make only a basic comparison with a singlet $s$-wave state. Within the statistical error of our data, the temperature dependent component of the induced moment $M_{\mathrm{spin}}$ is well described by a Yosida temperature dependence (see Fig.~\ref{Fig:TDep}) with $\Delta=1.78 k_B T=41$~K$=3.5$~meV.

\section{Conclusions}

We have used a polarised neutron diffraction technique to measure the induced magnetization density of near optimally doped Ba(Fe$_{0.935}$Co$_{0.065}$)$_2$As$_2$ ($T_C$=24~K) as a function of magnetic field  and temperature. The induced magnetization is confined to the Fe atoms with an oblate distribution spread out in the $a-b$ plane. The distribution is in reasonable agreement with a full potential LAPW band structure calculation which suggests that the relevant bands near the Fermi energy are of the $d_{xz/yz}$ and $d_{xy}$ type.

The $T$-dependence of the induced moment in the superconducting state is consistent with the Yosida function characteristic of spin-singlet pairing, and the induced moment is proportional to applied field for $\mu_0 H \leq \mbox{9 T} \approx \mu_0 H_{c2}/6$. We observe a large residual susceptibility $M (H \rightarrow 0,T \rightarrow 0)/H \approx 2/3 \chi_{\mathrm{normal}}$.  This is most easily interpreted as being due to the van Vleck contribution present in other $d$-band systems, but may also signal a residual quasiparticle density of states.

\section{Acknowledgements}
We thank P. J. Brown, A. Carrington, P. J. Hirschfeld and I. I. Mazin for helpful discussions.
Work at Stanford was supported by the Department of Energy, Office of Basic Energy Sciences under contract DE-AC02-76SF00515.
\bibliography{BaFeCoAs_D3}

\begin{thebibliography}{51}
\bibitem [{SF_()}]{SF_sign}%
  \BibitemOpen
  \href@noop {} {}\bibinfo {note} {The signs (phases) of $F_N(\mathbf{G})$ and
  $M(\mathbf{G})$ depend of the origin chosen for unit cell in
  Eq.~\ref{Eq:M_G}. The values in Table~\ref{table:D9} are used
  here.}\BibitemShut {Stop}%
\bibitem [{M_s()}]{M_sign}%
  \BibitemOpen
  \href@noop {} {}\bibinfo {note} {In this section we choose the unit cell
  origin to be at an Fe site. With this convention {$M(\mathbf{G}=002)$} is
  positive and corresponds to a positive susceptibility.}\BibitemShut {Stop}%
\end{thebibliography}
\bibliographystyle{apsrev4-1}

\end{document}